\documentclass[prd,aps,floats,twocolumn,nofootinbib]{revtex4-1}
\usepackage{slashed}
\usepackage{mathtools}
\usepackage{amsfonts}
\usepackage{amssymb}
\usepackage{epsfig}

\begin{document}

\newcommand{\m}[1]{\mathcal{#1}}
\newcommand{\nn}{\nonumber}
\newcommand{\ph}{\phantom}
\newcommand{\eps}{\epsilon}
\newcommand{\be}{\begin{equation}}
\newcommand{\ee}{\end{equation}}
\newcommand{\bea}{\begin{eqnarray}}
\newcommand{\eea}{\end{eqnarray}}
\newtheorem{conj}{Conjecture}

\newcommand{\plk}{\mathfrak{h}}
\newcommand{\bb}{\bar b}


\title{Overall signature of the metric and the cosmological constant}
\date{\today}


\author{Bruno Alexandre$^1$}
\author{Steffen Gielen$^2$}
\email{s.c.gielen@sheffield.ac.uk}
\author{Jo\~{a}o Magueijo$^1$}
\email{magueijo@ic.ac.uk}

\affiliation{$^1$Theoretical Physics Group, The Blackett Laboratory, Imperial College, Prince Consort Rd., London, SW7 2BZ, United Kingdom}
\affiliation{$^2$School of Mathematics and Statistics, University of Sheffield,
Hicks Building, Hounsfield Road, Sheffield S3 7RH, United Kingdom}

\begin{abstract}
We consider a little known aspect of signature change, where the overall sign of the metric is allowed to change, with physical implications. 
We show how, in different formulations of general relativity, this type of classical signature change across boundaries with a degenerate metric can be made consistent with a change in sign (and value) of the cosmological constant $\Lambda$. In particular, the separate ``mostly plus'' and ``mostly minus'' signature sectors of Lorentzian gravity are most naturally associated with different signs of $\Lambda$. We show how this general phenomenon allows for classical solutions where the open dS patch can arise from a portion of AdS space time. These can be interpreted as classical ``imaginary space'' extensions of the usual Lorentzian theory, with $a^2<0$.

\end{abstract}

\maketitle

\section{Introduction}

In classical gravity, it has been a convention to assume that the signature of the metric remains constant. However, this is not a requirement stipulated by the field equations themselves, but rather a prerequisite imposed on the metric prior to seeking solutions for Einstein's equations. By loosening this restriction, it becomes possible to discover solutions to the field equations that demonstrate a transition in signature or even a topology change \cite{Horowitz}, provided that they are appropriately interpreted, just as it happens in quantum cosmology \cite{HH1,Dereli}. Classical cosmological models describing metric signature change were introduced in \cite{Ellis1,Ellis2}, where it is shown that the singularity-free universes have an origin in time despite not having a true beginning. There are usually two approaches to address this problem, one considering a degenerate metric and the other a discontinuous metric \cite{Dray1,Dray2}. In the case of a degenerate metric, the metric signature changes as a result of a singularity. In contrast, a discontinuous metric changes signature through a topological transition. In both cases, the key is to identify specific conditions on the hypersurface that connects the
two disjoint regions with different signature. 

A possible relation between metric signature change and the cosmological constant problem \cite{Lombriser,Padilla,Weinberg} can be found in nonlinear first-order gravity \cite{Borow}. For instance, there is an interesting but somewhat ambiguous connection between the structure of the Lagrangian and the signature of the spacetime metric. By selecting specific functions of the non-linear Lagrangian, certain values of cosmological constants can be allowed, which correspond to Einstein metrics of varying signatures. It is even possible to choose a Lagrangian that produces coherent equations only for a particular signature.

Another model based on Friedmann--Robertson--Walker (FRW) metrics coupled to a self-interacting scalar field \cite{Ghaf} shows a relation between the cosmological constant and a transition from Euclidean to Lorentzian spacetime. It is found that there is an upper limit for the cosmological constant and that in proximity to the hypersurface where the signature changes, the scale factor and scalar field must remain constant. Furthermore, it was determined that solutions that involve a change in signature cannot exist when the scalar field is massless.

More recently, a signature changing metric and topology change have been used to ``save" unitarity for the Vilenkin wave function \cite{Bruno1}, with a model where a semi-infinite tower of Euclidean spheres is glued to half a de Sitter spacetime within a fundamentally real theory.

In this paper we study the more general case where, not only we can have change of signature, but also a ``flip'' of the metric, in which all metric components change sign simultaneously. 
This proposal for classical gravity goes beyond standard General Relativity where the metric 
cannot flip overall signature. The main point is that, while the overall metric signature is often presented as a pure convention, it can have real physical implications: the sign or absolute value of the cosmological constant can change across the surface where the metric is degenerate, and in general there is the freedom to make $\Lambda$ dependent on the flip parameter $\sigma$ (and also the signature $s$ distinguishing between Euclidean and Lorentzian). Thus, a transition in metric signature could be associated to a change in the energy density of dark energy, usually assumed to be fixed in standard cosmology.

To the best of our knowledge, this subtle interplay between $\Lambda$ and the overall metric signature has not been discussed directly at the classical level, although similar arguments feature in quantum cosmology. In particular, in \cite{HertogHartle} classical solutions including this flip appear as saddle-point representations of quantum cosmology; the flip  is used to relate the no-boundary proposal with positive $\Lambda$ 
to AdS/CFT (which reqiures $\Lambda<0$), or conversely negative $\Lambda$ theories to de Sitter solutions, by changing the overall signature. Our framework shows how to make such transitions consistent already classically and makes more explicit the choice one can make (which seems implicit in \cite{HertogHartle}).

To illustrate how signature change in gravity can be understood from different angles, we derive a minisuperspace model from three separate formalisms for general relativity. We start in Sec.~\ref{secmss} by writing a general action (including all possible signature cases) in the Einstein--Hilbert formalism and reducing it to minisuperspace. We introduce the variable $\tilde a^2$, which is an extension of the (normally positive-definite) squared scale factor $a^2$ to the whole real line. The surface where the flip occurs is defined in terms of the sign change of this new variable. There is also a new parametrisation of the cosmological constant in terms of a variable $\tilde\Lambda$, which likewise incorporates the effects of the flip. Proceeding to Sec.~\ref{secec}, we analyse the same problem in the first order formalism from the point of view of Einstein--Cartan theory. We find that the Bianchi identities do not require $\Lambda$ to be constant across surfaces with degenerate tetrad. A third approach towards introducing the same idea is from the perspective of the Pleba\'{n}ski formalism, introduced in Sec.~\ref{secplebanski}.
We discuss possible ``unimodular'' extensions of these theories in Sec.~\ref{secunimod}.
In Sec.~\ref{seceg} we focus on an example where one can observe topology change and a metric flip. For instance, we can transition from an open de-Sitter space to a flipped Lorentzian Anti-deSitter.

\section{Minisuperspace with arbitrary metric signature}\label{secmss}

We will be examining solutions where the signature changes across boundaries where the metric is degenerate. We write the metric signature generically as $\sigma (s,+,+,+)$, where we have introduced the ``flip'' parameter $\sigma=\pm 1$ (for reasons that will be clear presently) as well as $s=\mp 1$ for Lorentzian/Euclidean signature. The values of $s$ and $\sigma$ are determined by the eigenvalues of the metric, so we may think of them as functions $s(g_{\mu\nu})$ and $\sigma(g_{\mu\nu})$.  For Lorentzian signature, $\sigma=1$ corresponds to the ``East Coast convention'' whereas $\sigma=-1$ is the ``West Coast convention'', traditionally often favoured in particle physics. In Euclidean signature $\sigma=-1$ is the unusual convention of a negative definite metric.

Away from boundaries with degenerate metric, we take the action to be the 
Einstein--Hilbert action with a given value of the cosmological constant $\Lambda$:
\begin{equation}
S=\frac{1}{16\pi G}\int{\rm d}^4 x\,\sqrt{|g|}\,\left(\sigma R-2\Lambda\right)\,,
\label{eqEH}
\end{equation}
using conventions for $\sigma$ and $s$ that follow {\it most} of the literature. For $s=-1$, our convention amounts to requiring that $\Lambda>0$ leads to dS spacetime and $\Lambda<0$ to AdS, almost universally accepted. For $s=+1$ the situation is less clear: we follow~\cite{thieman} in not adding any $s$ dependence in the action, but other options exist (for example multiplying the $R$ term by $-s\sigma$). The $\Lambda$ term can also be multiplied by factors of $s$ and/or $\sigma$, leading to different conventions. We stress that in these conventions the action is always real, both in Euclidean and Lorentzian signature.

The Einstein equations resulting from (\ref{eqEH}) are
\begin{equation}
R_{\mu\nu}=\sigma\Lambda g_{\mu\nu}\,,
\end{equation}
and are invariant under a transformation $g_{\mu\nu}\rightarrow -g_{\mu\nu}$ which introduces two extra minus signs on the right-hand side while leaving the left-hand side invariant. Hence, the theory defined in this way and with $\Lambda>0$ has Lorentzian dS solutions of both ``East Coast'' or ``West Coast'' signature (and Euclidean spherical solutions with positive or negative definite metrics). Different conventions would lead to the possibility of both dS and AdS solutions within the same theory, as we will discuss below.

The spacetime Ricci scalar $R$ may be decomposed using the Codazzi equation
\begin{equation}
R = {\bf R}_3 - \sigma s \left(K_{ij}K^{ij}-K^2\right) + \nabla_\mu(\ldots)^\mu
\end{equation}
where ${\bf R}_3$ is the Ricci scalar of the spatial metric $h_{ij}$, $K_{ij}$ is the extrinsic curvature or second fundamental form, and we can cancel the total derivative $\nabla_\mu(\ldots)^\mu$ (whose detailed form is not important here) by adding an appropriate Gibbons--Hawking--York boundary term.

In a homogeneous isotropic (FRW) model, we may write the metric as
\begin{eqnarray}
{\rm d}s^2=\sigma s N^2\,{\rm d}t^2+\sigma a^2h_{ij}^0\, {\rm d}x^i\,{\rm d}x^j 
\label{metricparam}
\end{eqnarray}
where $N$ is a real lapse function and 
$h_{ij}^0$ is a fixed (positive definite) metric of constant curvature and $a$ is the scale factor, and
\begin{equation}
K_{ij}=\frac{1}{2N}\dot{h}_{ij}=\sigma\frac{a\dot{a}}{N}h_{ij}^0.
\end{equation}
The 3d Ricci scalar is ${\bf R}_3=6\sigma k/a^2$ in terms of a curvature parameter $k$. This defines $k$, which may again be seen as setting a convention (spheres always have $k>0$, regardless of metric signature). Finally, $\sqrt{|g|}=Na^3\sqrt{h^0}$.\footnote{This replacement, which is common in the literature, implicitly assumes $N\,a >0$. We could also write $\sqrt{|g|}=|Na^3|\sqrt{h^0}$, which would differ by a possible orientation factor ${\rm sgn}(N a)$. In this sense, the second order formulation makes a choice of ``tetrad orientation'' which makes it equivalent to first-order formalisms.} Putting it all together, (\ref{eqEH}) becomes 
\begin{equation}
S = \frac{3V_c}{8\pi G}\int {\rm d}t\;Na\left(k+s\frac{\dot{a}^2}{N^2}-\frac{\Lambda}{3} a^2\right),
\end{equation}
where $V_c=\int {\rm d}^3x\;\sqrt{h^0}$ is the coordinate volume of spatial slices.  We see that all $\sigma$ dependence cancels and $s$ only appears in a term with time derivatives, as one might have expected from a ``Wick rotation'' perspective. 

To bring this action into Hamiltonian form, notice that the conjugate momentum to $a^2$ is
\begin{equation}
p_a=\frac{\partial\mathcal{L}}{\partial((a^2)^\cdot)}=\frac{3V_c}{8\pi G}\frac{s\dot{a}}{N}
\end{equation}
and hence we may choose $\{b,a^2\}=8\pi G/3V_c$ with $b=-s\dot{a}/N$ as a conjugate pair. After another integration by parts, the Einstein--Hilbert action becomes
\begin{equation}
\label{MSSHamiltonian}
S=\frac{3V_c}{8\pi G}\int{\rm d}t\left(\dot{b}a^2-Na\left(sb^2-k+\frac{\Lambda}{3}a^2\right)\right)\,.
\end{equation}
The explicit form of matter Lagrangians coupled to this action will also be sensitive to spacetime signature. For example, for a scalar field we would expect the kinetic term to change sign when going to Euclidean signature.

In this derivation we have tacitly assumed $a^2>0$ and $N^2>0$, but nothing in the final result (\ref{MSSHamiltonian}) depends on the additional sign $\sigma$ that we included. Alternatively, we could absorb $\sigma$ into a more general definition of $a^2$ in which it can be positive or negative. We can define 
\begin{equation}
    a^2=\sigma \tilde a^2,
    \label{eqat}
\end{equation}
with $\sigma={\rm sgn}(\tilde a^2)$, so that $-\infty< \tilde a^2<\infty $, trading $a^2$ for a new $\tilde a^2$ covering the whole real line. We also redefine the lapse function as
\begin{equation}
  \tilde{N}^2=\sigma N^2.
\end{equation} 
Then, in terms of the new variables, the metric is
\begin{equation}
    {\rm d}s^2 = s\tilde{N}^2\,{\rm d}t^2 + \tilde a^2\,h_{ij}^0{\rm d}x^i{\rm d}x^j\,
\end{equation}
which might be seen as a more economical parametrisation than (\ref{metricparam}), 
and (\ref{MSSHamiltonian}) becomes
\begin{equation}
S=\frac{3V_c}{8\pi G}\int{\rm d}t\left(\dot{\tilde{b}}\tilde a^2-\tilde{N}\tilde{a}\left(s\tilde b^2-k+\frac{\tilde{\Lambda}}{3}\tilde a^2\right)\right)
\label{eqmsst}
\end{equation}
with $\tilde{b}=\sigma b$ and $\tilde{\Lambda}=\sigma \Lambda$.
Hamilton's equations are
\begin{align}
    \dot{\tilde{b}} &= \tilde{N}\tilde{a}
    \frac{\tilde{\Lambda}}{3}\,,
   \\ ({\tilde a}^2)^\cdot&=-2\tilde{N}\tilde{a}\,s\tilde b.\label{a2HamEq}
\end{align}
 For $\tilde{N}^2<0$ ($\tilde{a}^2<0$), the quantity $\tilde{N}$ ($\tilde{a}$) requires a choice of complex square root (see Sec.~\ref{seceg}); we have made a choice in replacing $N a$ by $\tilde{N} \tilde{a}$ (rather than $-\tilde{N} \tilde{a}$). Moreover, $\tilde{\Lambda}$ is no longer a fixed constant but changes sign according to
\begin{equation}
 \tilde{\Lambda}={\rm sgn}(\tilde a^2)\Lambda\,.
\end{equation}
In terms of $\tilde\Lambda$, Lorentzian dS in ``mostly plus'' signature has $\tilde\Lambda>0$ but in ``mostly minus'' signature has $\tilde\Lambda<0$. This simply reflects our chosen assumptions for how $\Lambda$ changes with changing signature. Had we started with the opposite convention, i.e., $\Lambda>0$ with ``mostly minus'' signature gives Lorentzian AdS, $\tilde\Lambda$ would be constant. 

In general, $\tilde\Lambda$ is locally constant in each region in which the metric is non-degenerate, but can flip sign in passing through a boundary of degenerate metric. We can use this freedom to choose $\Lambda$ as a general function of the signature parameters $\sigma$ or $s$,
\begin{equation}
    \Lambda=\Lambda(\sigma,s)\,. 
\end{equation}
Through the contracted Bianchi identities, the Einstein equations imply
\begin{equation}
\partial_\mu(\sigma\Lambda) = 0
\end{equation}
whenever the Einstein equations are well-defined. At the boundaries with degenerate metric, curvature tensors are not defined and there are no Einstein equations imposing such a restriction.

At the level of minisuperspace (\ref{eqmsst}), the consistency condition on $\tilde\Lambda$ is obtained by taking the time derivative of the Hamiltonian constraint and using the equations of motion:
\begin{equation}
  \dot{\tilde\Lambda}\,\tilde{a}^2=0\,.
  \label{consistency}
\end{equation}
Again this shows that $\tilde\Lambda$ needs to be constant away from degenerate points $\tilde{a}^2=0$.

The prescription of constant $\tilde{\Lambda}$, which is the one we will use below and the one also appearing in the quantum cosmology literature \cite{HertogHartle}, is evidently compatible with the Bianchi identities in any case.

\section{Signature change in Einstein--Cartan formalism}\label{secec}
To understand better the implications of the previous Section, we now consider how it fits into the first order formulation (in whatever guise) which allows for topology and signature change even at the  classical level~\cite{Horowitz}. For definiteness we take the Einstein--Cartan formalism. The space-time metric is defined in terms of the tetrad $e^a_{\; \mu}$ by $g_{\mu\nu}=\eta_{ab}e^a_{\;\mu} e^b_{\; \nu}$, where by convention we put the signature $s$ and flip $\sigma$ into the tangent-space metric 
$\eta_{ab}={\rm diag}[\sigma(s+++)]$ rather than the tetrad. 
The Levi-Civita symbol remains the same for $\epsilon_{abcd}$, but versions with raised indices need to be adapted accordingly.  
Then, the action for gravity with a cosmological constant is
\begin{eqnarray}\label{ECaction}
S_{G}&=&\frac{1}{32\pi G}\int  \epsilon_{abcd}\left( \sigma e^a e^b R^{cd} -\frac{\Lambda}{6} e^a e^be^ce^d\right)
\end{eqnarray} 
where $R^{ab}$ is the curvature 2-form associated to a connection  ${\Gamma^a}_b$.
Under the usual conditions (zero torsion and non-degeneracy of the metric) this action reduces to (\ref{eqEH}). 

The flip $\sigma$ 
remains a notational nuisance without any physical effect until we realize that in the first-order formalism the signature can change across surfaces with degenerate metric. In tandem with this, $\Lambda$ (as defined by any convention, e.g.~(\ref{eqEH}) or~(\ref{ECaction})) can
change sign and value across such  boundaries, indeed it can be any function $\Lambda=\Lambda(\sigma,s)$. Each of these functions leads to physically different theories. 
To see how this is possible, consider the equations of motion following from (\ref{ECaction}),
\bea
\epsilon_{abcd}{\left(e^b  R^{cd}-\frac{\sigma  }{3}\Lambda e^b  e^c  e^d\right)}=0\,,\label{Eeq}\\
T^{[a}  e^{b]}=0\,.
\eea
Usually the Bianchi identity $DR=0$ forces $\Lambda$ to be constant; however, (setting the torsion to zero; see~\cite{Lee1,Lee2,TomZ}) we have
\begin{equation}\label{bianchi}
    \epsilon_{abcd} e^b e^c e^d {\rm d} (\sigma\Lambda)=0
\end{equation}
so that $\sigma \Lambda$ can vary across surfaces where the tetrad is degenerate, such as a 3-surface where at least one of the tangent tetrads is zero, as is often the case in (not necessarily isotropic) cosmology. Again, any prescription $\Lambda=\Lambda(\sigma,s)$ seems consistent. Since $s$ and $\sigma$ are scalars this does not break diffeomorphism invariance. 

This may make some conventions more adapted to some theories, should the only dependence of $\Lambda$ on $s$ and $\sigma$ be a signal.  For example (\ref{ECaction}) is best for theories in which $\Lambda$ defined in this way is indeed a constant. We will presently consider a theory where
\begin{equation}
\tilde \Lambda = \sigma \Lambda
\end{equation}
is the variable kept constant. For such a theory it is more suggestive to use the notation
\begin{eqnarray}
S_{G}&=&\frac{1}{32\pi G}\int  \sigma \epsilon_{abcd}\left( e^a e^b R^{cd} -\frac{\tilde\Lambda}{6} e^a e^be^ce^d\right)\,. 
\label{}
\end{eqnarray} 
In such a first-order theory there are solutions where, for example, portions of dS can be glued to portions of AdS. We will make this explict later in this paper, but the process is not dissimilar to the constructions in~\cite{Horowitz}.

We can reduce the Einstein--Cartan action (with generalized flip and signature)  to homogeneous and isotropic solutions by making the ansatz $\Gamma^i_{\;0}=-s\frac{b}{a} e^i$ and the usual ones for the tetrad and the other components of the connection. Then (\ref{ECaction}) reduces to
\begin{equation}
\label{MSSaction}
S=\frac{3V_c}{8\pi G}\int{\rm d}t\left(\dot{b}a^2- Na\left(s b^2- k+  \frac{\Lambda(\sigma,s)}{3}a^2\right)\right)\,,
\end{equation}
where we explicitly introduce the dependence of $\Lambda$ on $s$ and $\sigma$. This is now exactly (\ref{MSSHamiltonian}), obtained here without having to choose an explicit prescription for $\sqrt{|g|}$.

Note that, with the chosen ansatz for  $\Gamma^i_{\;0}$,  the torsion-free condition $T^i=De^i=0$ (which is an equation of motion) implies
\begin{equation}
    b = - s \frac{\dot a}{N}\,,
\end{equation}
consistent with Hamilton's equation for $a^2$ (cf.~(\ref{a2HamEq})). We could of course have chosen a different ansatz for $\Gamma^i_{\;0}$ (or definition of $b$), making $s$ appear in the Poisson bracket but disappear in the on-shell expression for $b$ or $\dot a$. 

In the above we have included different signatures, but, just as with the Einstein--Hilbert formulation, we have tacitly assumed $a^2\ge 0$ (and that all variables and parameters are real). We could also fold the signature information in $\sigma$ into a more general definition of $a$ and $N$ within a theory with nominally fixed signature\footnote{We can play the same trick with $b^2$ and $s$, but will leave $s$ fixed here, for clarity.}. We could define again $a^2$ by (\ref{eqat}), but this time defining 
\begin{eqnarray}
    \sigma={\rm sgn}(\tilde a^2),
\end{eqnarray}
so that  (\ref{MSSaction}) becomes (\ref{eqmsst}). We now have a theory with fixed $\sigma$ where the opposite $\sigma$ is represented by imaginary $a$ and $N$ and real $b$ (where $s$ was left unchanged). The action itself remains real.

If $\Lambda$ is kept constant (if $\Tilde{\Lambda}\rightarrow - \Tilde{\Lambda}$ when $\tilde a^2\rightarrow -\tilde a^2$) we have perfectly identical solutions in the $a^2<0$ regions to those in the $a^2>0$. They are physically indistinguishable and reflect nothing but different $\sigma$ conventions. Indeed, whatever the actual dependence of $\tilde \Lambda$ on $\sigma$, $\Tilde{\Lambda}\rightarrow - \Tilde{\Lambda}$ and $\tilde a^2\rightarrow -\tilde a^2$ remains a symmetry of (\ref{eqmsst}).

But if instead $\Tilde{\Lambda}$ is kept fixed physically different solutions arise.  We will investigate these solutions in Sec.~\ref{seceg}.

\section{Signature change in Pleba\'{n}ski formulation}
\label{secplebanski}

Another way of looking at general relativity is the (chiral) Pleba\'{n}ski formulation which encodes the metric into complex self-dual 2-forms, and gravity is mediated by a complex $SU(2)$ connection \cite{Plebanski,PracticalIntro,KirillBook}. An action for general relativity with ``cosmological constant'' $\tilde\Lambda$ is given by
\begin{equation}S=\frac{1}{\sqrt{s}\,8\pi G}\int \Sigma^i F_i -\frac{1}{2}M_{ij}\Sigma^i\Sigma^j+\frac{1}{2}\nu({\rm tr}\,M-\tilde\Lambda)
\end{equation}
where $F^i$ are the curvature 2-forms of a connection $A^i$, $\Sigma^i$ are 2-forms and $M_{ij}$ is a symmetric $3\times 3$ matrix. The prefactor $\sqrt{s}$ ensures that the action is real for both Lorentzian and Euclidean solutions.

The equations of motion are
\begin{align}
D_A\Sigma^i &=0\,, \label{PlebEq1} \\ F^i &=M^{ij}\Sigma_j\,, \\ \Sigma^i\Sigma^j & = \delta^{ij}\nu \label{PlebEq2}
\end{align}
together with the constraint ${\rm tr}\, M=\tilde\Lambda$ coming from variation with respect to $\nu$. Solutions to these equations represent solutions of complex general relativity. Lorentzian solutions are obtained by imposing reality conditions
\begin{equation}
{\rm Re}(\Sigma^i\Sigma_i)=\Sigma^i\bar{\Sigma}^j=0
\end{equation}
whereas Euclidean solutions correspond to real $\Sigma^i$. 

A metric can be recovered from the $\Sigma^i$ using the Urbantke formula \cite{Urbantke}
\begin{align}
g_{\mu\nu}\,\varepsilon_{\Sigma}&=-\frac{\sqrt{s}}{6}\epsilon_{ijk}\,i_\mu\Sigma^i\,i_\nu\Sigma^j\,\Sigma^k\,,
\\\varepsilon_{\Sigma}&=\frac{\sqrt{s}}{6}\Sigma^i\Sigma_i
\end{align}
where $i_\mu\Sigma^i$ is a 1-form defined by $(i_\mu\Sigma^i)_\nu=\Sigma^i_{\mu\nu}$. The Urbantke metric is Lorentzian (Euclidean) for solutions satisfying reality conditions, but its signature can never be fixed: for any solution to (\ref{PlebEq1})-(\ref{PlebEq2}), $\Sigma^i\rightarrow -\Sigma^i$ and $M_{ij}\rightarrow -M_{ij}$
give another solution whose Urbantke metric has opposite signature, with the sign of $\tilde\Lambda$ also reversed. This is where the ``flip'' $\sigma$ appears in the Pleba\'{n}ski formalism. If the field equations are satisfied, the Urbantke metric $g_{\mu\nu}$ satisfies $R_{\mu\nu}=\tilde\Lambda g_{\mu\nu}$, which shows that the parameter $\tilde\Lambda$ corresponds to $\tilde\Lambda=\sigma\Lambda$ defined in the metric formulation. Of course, nothing would stop us from replacing $\tilde\Lambda$ by a general function $\Lambda(\sigma,s)$.\footnote{Again, there is a Bianchi identity which requires $\tilde\Lambda$ to be constant whenever the Urbantke metric is non-degenerate \cite{ElliotSteffen2}.}

A more compact form of the Pleba\'{n}ski action is obtained by integrating out the $\Sigma^i$ via $\Sigma^i=(M^{-1})^{ij}F_j$, which yields
\begin{equation}S=\frac{1}{\sqrt{s}\,16\pi G}\int (M^{-1})_{ij}F^i F^j+\nu({\rm tr}\,M-\tilde\Lambda)\,.
\end{equation}
The metric is now implicitly encoded in the matrix $M$.

We can reduce this action to minisuperspace via \cite{ElliotSteffen}
\begin{align}
A^i&=(\sqrt{s}b+\sqrt{k})e^i\,,\\ M_{ij}&=\frac{k-sb^2}{p}\delta_{ij}
\end{align}
where $e^i$ is a suitable fiducial triad. The action reduces to
\begin{equation}
S = \frac{3V_c}{8\pi G}\int p\dot{b}-M\left(\tilde\Lambda-\frac{3(k-sb^2)}{p}\right)
\end{equation}
with a new Lagrange multiplier $M$. Identifying $Na:=3M/p$ and $a^2:=p$ now reproduces exactly the minisuperspace action (\ref{eqmsst}) that we already found in Einstein--Hilbert or Einstein--Cartan theory. From the perspective of Pleba\'{n}ski gravity the sign of $p$ (or $a^2$) is undefined {\em a priori}, without any need to introduce a $\sigma$ explicitly, or apply the redefintions needed in the other formalisms. Classical signature change is not optional but mandatory here. The fact that the same theory (with fixed but arbitrary non-zero $\tilde\Lambda$) always contains both adS and dS solutions was observed in \cite{ElliotSteffen}.

\section{Unimodular extension}
\label{secunimod}

One may feel that the indeterminacy of the conserved function $\tilde\Lambda=
\tilde\Lambda(\Lambda,\sigma,s)$ is a weakness in all of the different formalisms discussed here. This may be softened using a formalism analogous to the unimodular extension of gravity due to Henneaux and Teitelboim \cite{HenneauxTeitelboim}, where the cosmological constant at any rate appears as a constant of motion rather than a fixed parameter in the action. For example we could add to $S_G$ a new term containing the vector density $\cal T^\mu$:
\begin{equation}
    S_G\rightarrow S_G - \int d^4x\, \tilde \Lambda \partial_\mu {\cal T}^\mu
\end{equation}
Note that the unimodular term does not contain the  metric, so it is still well defined where the metric is degenerate. Hence it can indeed be used
to define the ``input” as part of the action. The equations of motion would then say that $\partial_\mu\tilde\Lambda=0$, regardless of what $\sigma$ and $s$ are doing, but the value of $\tilde\Lambda$ is not specified and can vary between classical solutions. In this sense, such a unimodular extension is both more specific (it forces $\tilde\Lambda$ to be constant rather than an arbitrary function of $\sigma$ and $s$) and more general, since the value $\tilde\Lambda$ cannot be prescribed.

\begin{figure}[ht]
\begin{tabular}{c c}
\includegraphics[scale=0.5]{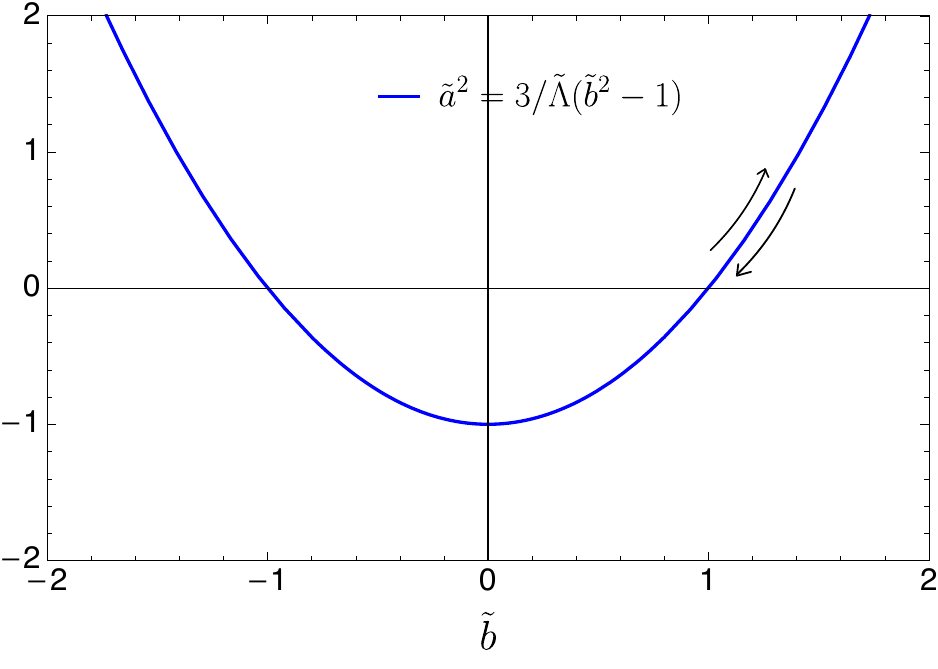}
\end{tabular}
\caption{The on-shell (Hamiltonian) condition (\ref{HamOpenI}) phrased in terms of $\tilde a^2$ as a function of $b$ for $k=-1$, $\tilde\Lambda>0$. The arrows indicate a bounce in $\tilde b$ corresponding to the standard expanding $k=-1$ dS patch, with only $\tilde a^2>0$. (The same holds true at $\tilde b=-1$, for the corresponding contracting solution.)} 
\label{figa2ref}
\end{figure}

\section{An example: open inflation from AdS beginnings}\label{seceg}

The case $\tilde \Lambda>0$ and $k=-1$ (open Universe) with $s=-1$ (Lorentzian signature),  also known as  ``open inflation'', 
is interesting because it contains classical solutions that change $\sigma$ at $\tilde a=0$, where the metric is degenerate but there is no curvature singularity. This is manifest from the Hamiltonian constraint
\begin{equation}\label{HamOpenI}
    \tilde b^2-1=\frac{\tilde \Lambda}{3}    
    \tilde a^2\,,
\end{equation}
with a constant positive $\tilde \Lambda$. 
As we see in Figure~\ref{figa2ref}, the mass-shell condition includes a region with $\tilde a^2<0$.  
If we impose $\tilde a^2>0$ (no flip $\sigma$ change) this leads to the usual two branches (contracting and expanding)
\begin{eqnarray}
    \tilde a&=&\pm \sqrt{\frac{3}{\tilde\Lambda}} \sinh \sqrt{\frac{\tilde \Lambda }{3 }}t\,, \\
    \tilde b&=& \pm \cosh \sqrt{\frac{\Tilde{\Lambda} }{3 }} t
\end{eqnarray}
in the $N=1$ gauge. Then, just as $|\tilde b|$ is about to drop below 1, requiring access to $\tilde a^2<0$, there is a ``bounce'' in $\tilde b$ at $\tilde b=1$ ($\tilde b=-1$) and $t=0$, with $\tilde b\ge 1$ ($\tilde{b}\leq -1$) and so $\tilde a^2>0$ always holding true. This corresponds to the expanding (contracting) open dS solution as we have $\tilde \Lambda>0$ and $\sigma=1$, which means $\Lambda>0$. Such a situation is schematically represented by an arrow in Fig.~\ref{figa2ref} for the expanding branch (the mirror image would be the contracting branch).

But (\ref{HamOpenI}) is also consistent with $-1<\tilde b<1$ and $-3/{\Tilde{\Lambda}}<\tilde  a^2<0$. A few adjustments are then in order. The Hamilton's equations are the usual
\begin{eqnarray}
    \dot {\tilde a}&=&\tilde N \tilde{b}\,,\label{Ham1}\\
    \dot {\tilde b}&=&\frac{\tilde \Lambda}{3} \Tilde{N}\Tilde{a}\,, \label{Ham2}
\end{eqnarray}
so we need $\tilde N$ to be imaginary where $\tilde a$ is imaginary to keep $\tilde b$ real (and so $s=-1$, i.e., the theory is Lorentzian).
In the $\tilde N=i$ gauge the solutions to (\ref{HamOpenI}), (\ref{Ham1}) and (\ref{Ham2}) are
\begin{eqnarray}
    \tilde a&=&\pm \sqrt{\frac{3}{\tilde\Lambda}}  i \sin \sqrt{\frac{\tilde \Lambda }{3 }}t\,, \\
    \tilde b&=& \pm \cos \sqrt{\frac{\Tilde{\Lambda} }{3 }} t\,. 
\end{eqnarray}
This is physically equivalent to the cosmological half-diamond of AdS since $\tilde\Lambda>0$ but now we have flipped signature, $\sigma=-1$, implying $\Lambda<0$.

Since the metric is degenerate at $a=0$, we can glue the solutions with $\tilde a^2>0$ to the ones with $\tilde a^2<0$ in the first order formalism; such gluing is consistent with Bianchi identities as we showed in the previous sections. As a matter of fact, although here we are keeping $\Tilde{\Lambda}$ constant, any function $\Tilde \Lambda=\Tilde \Lambda(\sigma)=\tilde\Lambda({\rm sgn}(\tilde a^2))$ would be consistent, due to (\ref{consistency}). 

\begin{figure}[ht]
\begin{tabular}{c c}
\includegraphics[scale=0.5]{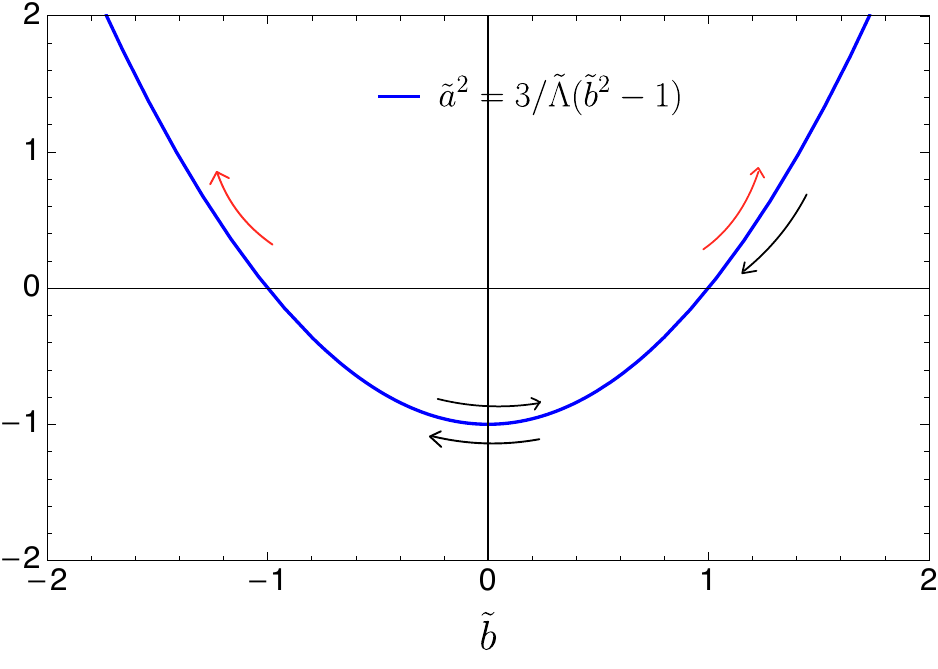}
\end{tabular}
\caption{The on-shell (Hamiltonian) condition (\ref{HamOpenI}) phrased in terms of $\tilde a^2$ as a function of $b$ for $k=-1$, $\tilde\Lambda>0$. The arrows indicate the connection between the dS and AdS space: an incident wave from $\tilde a^2>0$ at $\tilde b=1$ is transmitted to the region $\tilde a^2<0$, where it can be cycled indefinitely, and eventually get back to $\tilde a^2>0$ at $\tilde b=\pm 1$.} 
\label{figa2tran}
\end{figure}

We can therefore glue the two types of solution related  by a flip change found above. Indeed, we can glue the open de Sitter solution to a flipped Lorentzian portion of AdS executing as many cycles as we want and then connecting with the $\Tilde{a}^2>0$, $\tilde b\le -1$ solution, or even back to the $\tilde a^2>0$, $\tilde b\ge 1$. This is represented in Fig.~\ref{figa2tran}.  

\begin{figure*}[ht]
\centering
\begin{tabular}{c c}
\includegraphics[scale=0.3]{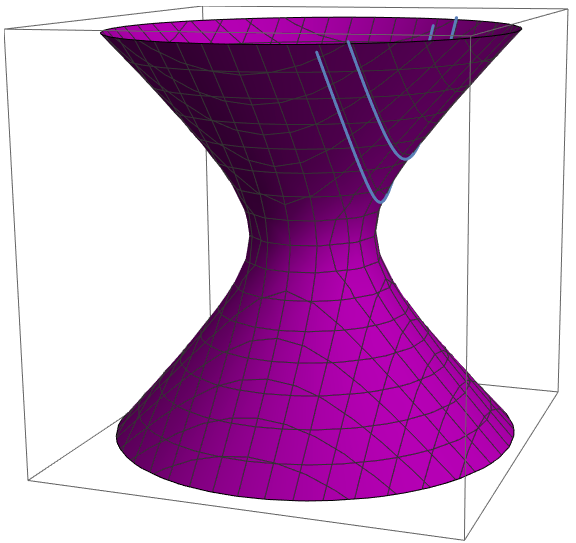} & \includegraphics[scale=0.3]{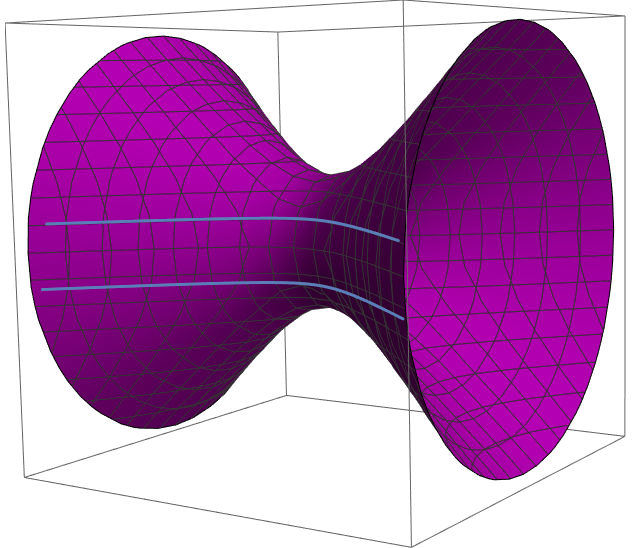} \\
\includegraphics[scale=0.3]{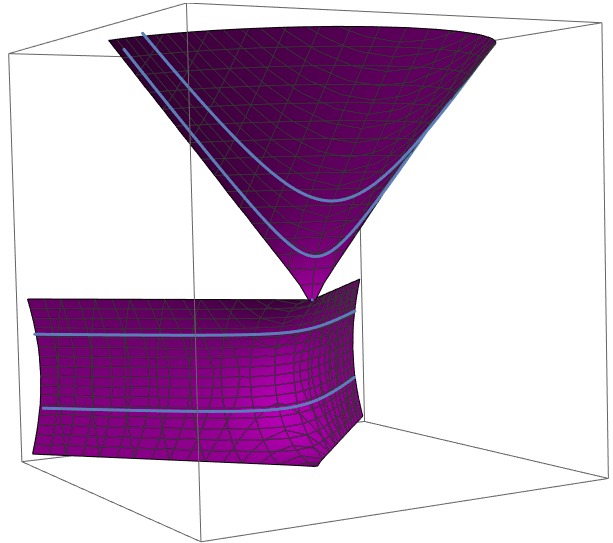}  & 
\includegraphics[scale=0.3]{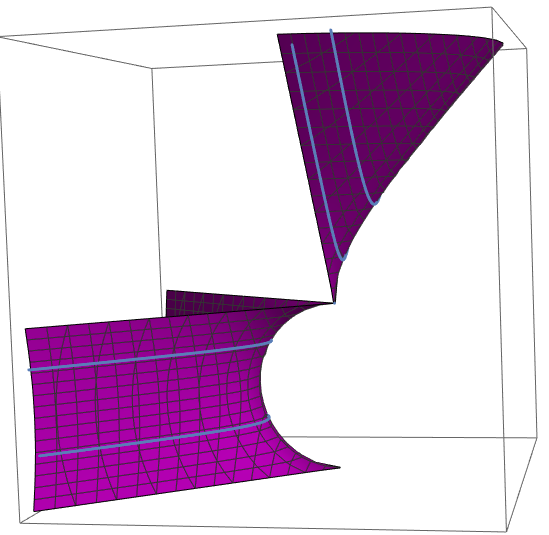}
\end{tabular}
\caption{Top panel: On the left, the cosmological patch (blue lines) on the de Sitter spacetime for $k=-1$ corresponding to slices of constant time $t>0$. On the right, we have the same description but instead for the Anti-de Sitter spacetime and for $t<0$. Bottom panel: glued version of the universe where we connect dS with AdS at $t=0$ where we have a flip in the metric signature at $a=0$. } 
\label{figuni}
\end{figure*}

In Fig.~\ref{figuni}, we represent this glued version of spacetime in a $3$D plot displaying the cosmological patches in both spaces.
 
\section{Conclusion}

In this paper we have shown how the metric can "flip" signature across a hypersurface characterised by a degenerate tetrad. Additionally, the cosmological constant is allowed to change across these surfaces (since the Bianchi identities fail to force its constancy in that context). 
We used three different approaches to this problem: Einstein--Hilbert, Einstein--Cartan and Plebański gravity. The point was then to realise the metric flip in these three different cases.

In the EH formulation one usually finds that the contracted Bianchi identities and the Einstein equations imply the constancy of 
$\Lambda$. When the metric is degenerate, the field equations are no longer well defined, so $\Lambda$ can vary across the boundary as a function of both $\sigma$ and $s$. In contrast, in the EC formalism, the field equations are still well defined on the boundary where there is a signature flip, but the tetrad is degenerate and the Bianchi identities fail to impose the constancy of  $\Lambda$. Finally, in Plebański gravity, the flip happens through the Urbantke metric, whose signature is never fixed.

In all three cases we reduced the respective actions to minisuperspace: the same result is obtained (which is unsurprising, as most degrees of freedom are frozen). We then investigated the implications of these results in the isotropic and homogeneous case with a concrete example: ``open inflation'', where we found a transition from an open dS to a flipped open patch of AdS spacetime. The  connecting point is characterized by $a^2=0$ where the metric flips signature. In this specific example, we have chosen the setting in which the variable $\tilde\Lambda$ is kept fixed and positive, so that $\Lambda$ acquires a dependence on  $\sigma$. Other choices are possible and may lead to a range of interesting examples. The dependence of $\Lambda$ on the signature parameter $s$ may be of interest in Euclidean/Lorentzian transitions,  see, e.g., \cite{Kothawala:2017nva} for somewhat related ideas.

In the context of gauge based approaches to quantum gravity it should not be surprising that $a^2<0$ is a possibility. Indeed $a^2$ (and not $a$) is the minisuperspace reduction of the densitized inverse triad, which plays the role of the electric field, conjugate to the gauge field. There is nothing exuberant about negative  electric fields. This is forced upon us in the quantum theory~\cite{JoaoPaper,Steffen,bruno2} (see also \cite{ElliotSteffen}), should we require unitarity and starting from the connection representation. It is interesting that even classically we can find new solutions exhibiting what presumably is the peak of the wave packets moving from the $a^2<0$ to the $a^2>0$ region.

\acknowledgments
This work was supported by the FCT Grant No. 2021.05694.BD (B.A.), the Royal Society through the University Research Fellowship Renewal URF$\backslash$R$\backslash$221005 (S.G.) 
and the STFC Consolidated Grant ST/T000791/1 (J.M.). 


\end{document}